\documentstyle[12pt]{article}
\title{\bf Exact solutions of Dirac equation on a 2D gravitational background}
\author{S. K. Moayedi$^{a, c}$,
F. Darabi$^{b, c}$\thanks{Corresponding author, e-mail:
f.darabi@azaruniv.edu}\\ $^{a}${\small Department of Physics, Arak
University, Arak, Iran.}\\ $^{b}${\small Department of Physics,
Azarbaijan University of Tarbiat Moallem , Tabriz,  Iran.}\\
$^{c}${\small Research Institute for Fundamental Sciences, Tabriz,
Iran. }}
\begin{document}
\maketitle \baselineskip.4in
\begin{abstract}
We obtain classes of two dimensional static Lorentzian manifolds,
which through the supersymmetric formalism of quantum mechanics
admit the exact solvability of Dirac equation on these curved
backgrounds. Specially in the case of a modified supersymmetric
harmonic oscillator the wave function and energy spectrum of Dirac
equation is given explicitly.
\end{abstract}
\newpage
\section*{I Introduction}

To describe the physics governing the dynamics of scalar and
spinor particles the Klein-Gordon and Dirac equations must be
solved. In general, solving these equations in (3+1)-dimensional
curved background is difficult and a weak field approximation
\cite{1} may be required, or asymptotic solutions may be obtained
\cite{2}. One can also solve these equations by numerical methods
\cite{3}, and WKB approximations \cite{4}. An alternative approach
is to consider lower-dimensional space-times and obtain exact
solutions \cite{5}. This may help us to get a deeper insight into
general features of (3+1)-dimensional problems.

In a previous work \cite{8'}, we have solved exactly the
Klein-Gordon equation on a static 2-dimensional space-time, by
using the standard techniques of supersymmetric quantum mechanics.
The purpose of the present paper is to solve exactly the Dirac
equation on a 2-dimensional conformally flat static space-time
and then find its energy spectrum. We are interested in solving the Dirac
equation through the supersymmetric formalism of quantum mechanics
\cite{5', 6'}.

\section*{II Dirac equation on a (1+1)-dimensional Lorentzian manifold }

Dirac's equation in curved space-time requires the use of {\it
bein} formalism to project the spinors into a Minkowskian local
inertial frame \cite{1'}. In our notation, the Latin indices refer
to local inertial frame with the metric $\eta_{ab}$, while the
Greek ones refer to curved space-time with the metric $g_{\mu
\nu}$. The zweibeins $e_{\: \mu}^{a}$ are used to project the
vectors between the two frames, and satisfy the following relation
\cite{2'}
\begin{equation}
g_{\mu \nu}(X)=\eta_{ab} e_{\:\mu}^{a}(X) e_{\:\nu}^b (X),
\label{1}\end{equation} where $X:=(t, x)$, and
$\eta_{ab}=\mbox{diag}(+1, -1)$. In flat space-time the Dirac
equation is written as
\begin{equation}
i \gamma^a \partial_a \Psi-m\Psi=0,
\label{99}
\end{equation}
where the $\gamma$ matrix conventions are
$$
\{ \gamma^a, \gamma^b \}=2 \eta^{ab} \:\:,\:\:
\sigma^{ab}=\frac{1}{4}[ \gamma^a, \gamma^b ], \label{7}
$$
with $\sigma^{ab}$ being the generator of Lorentz rotations.

Dirac equation is generalized to curved space-time through the
spin connections $\omega^{bc}_{\:\: \mu}$
\begin{equation}
i \gamma^{\mu} \nabla_{\mu} \Psi-m\Psi=0,
\label{98}
\end{equation}
where
$$
\nabla_{\mu}=\partial_{\mu}+\frac{1}{2}\sigma^{bc}\omega_{bc \mu}
\:\:\:,\:\:\: \omega_{bc \mu} \equiv E_b^{\: \nu} E_{c \nu};
$$
with the semicolon denoting covariant differentiation, and
$$
\gamma^{\mu}=\gamma^a E_a^{\: \mu} \:\:\:,\:\:\: \{ \gamma^{\mu},
\gamma^{\nu} \}=2g^{\mu \nu},
$$
with $E_a^{\: \mu}$ being the inverse of $e_{\: \mu}^{a}$.

Now, we take the static conformally flat metric $g_{\mu \nu}$
\begin{equation}
ds^2=e^{\sigma(x)} (dt^2-dx^2),
\label{10}
\end{equation}
where $\sigma(x)$ is a dilatonic static field. Using the relation
(\ref{1}), the zweibeins and their inverse corresponding to the
metric (\ref{10}) are respectively obtained
\begin{equation}
e^a_{\: \mu}=\left(\begin{array}{cc} e^{\frac{1}{2}\sigma(x)} & 0
\\ 0  & e^{\frac{1}{2}\sigma(x)}
\end{array}\right)_,
\label{11}
\end{equation}
\begin{equation}
E^{\: \mu}_{a}=\left(\begin{array}{cc} e^{-\frac{1}{2}\sigma(x)} &
0
\\ 0  & e^{-\frac{1}{2}\sigma(x)}
\end{array}\right)_.
\label{12}
\end{equation}
It is well known that Dirac equation in conformally flat
space-time is identical to the Minkowskian one ( apart from a
conformal factor ) once an appropriate transformation is employed
on the spinor $\Psi$.

In this regards, one may obtain the (1+1)-dimensional Dirac
equation in the curved space-time (\ref{10}) by employing the
following transformations
$$
\gamma^a \partial_a \rightarrow e^{-\frac{\sigma}{2}} \gamma^a
\partial_a,
$$
$$
\Psi \rightarrow e^{\frac{\sigma}{4}} \Psi,
$$
on the Dirac equation (\ref{99}). Therefore, we obtain
\begin{equation}
[(i\gamma^0 \partial_t +i\gamma^1
\partial_x)+\frac{i}{4}\gamma^1\frac{d\sigma(x)}{dx}-me^{\frac{1}{2}\sigma(x)}]\Psi(X)=0,
\label{13}
\end{equation}
where we consider $\Psi$ as
\begin{equation}
\Psi(X)= \left(\begin{array}{c} \Psi_1(X) \\ \Psi_2(X)
\end{array}\right)_.
\label{9}
\end{equation}
Following Jackiw and Rebbi \cite{4', 5'}, we take the following
representations for the $\gamma^a$ matrices
\begin{equation}
\gamma^0=\sigma^1\:\:\:,\:\:\:\gamma^1=i\sigma^3,
\label{8}
\end{equation}
where $\sigma^1, \sigma^3$ are the Pauli
spin matrices. If we now operate $\gamma^0$, from left, on both
sides of Eq. (\ref{13}), we have
\begin{equation}
\left(\begin{array}{cc} 0 &
-\partial_x-\frac{1}{4}\frac{d\sigma(x)}{dx}+me^{\frac{1}{2}\sigma(x)}
\\
\\ \partial_x+\frac{1}{4}\frac{d\sigma(x)}{dx}+me^{\frac{1}{2}\sigma(x)}  &
0
\end{array}\right)\Psi(X)=i\partial_t \Psi(X). \label{14}
\end{equation}

\section*{III Exact solutions of 2D Dirac equation by SUSY QM methods}

In this section, we will find the exact solutions of Dirac
equation by using the standard techniques of supersymmetric
quantum mechanics. By assuming the time dependence of the two
component spinor $\Psi(X)$ as
\begin{equation} \Psi(X)=e^{-i{\cal E}t} \psi(x), \label{15}
\end{equation}
Eq.(\ref{14}) can be written as the following eigenvalue equation
\begin{equation}
H_D \psi(x)={\cal E}\psi(x), \label{16}
\end{equation}
where $H_D$, and $\psi(x)$ are
\begin{equation}
H_D=\left(\begin{array}{cc} 0 &
-\frac{d}{dx}-\frac{1}{4}\frac{d\sigma(x)}{dx}+me^{\frac{1}{2}\sigma(x)}
\\
\\ \frac{d}{dx}+\frac{1}{4}\frac{d\sigma(x)}{dx}+me^{\frac{1}{2}\sigma(x)}  &
0
\end{array}\right)_, \label{17}
\end{equation}
\begin{equation}
\psi(x)= \left(\begin{array}{c} \psi_1(x) \\ \psi_2(x)
\end{array}\right)_.\label{18}
\end{equation}
By defining the transformation matrix
\begin{equation}
U(x)=\left(\begin{array}{cc} e^{\frac{1}{4}\sigma(x)} & 0
\\
\\ 0 & e^{\frac{1}{4}\sigma(x)}
\end{array}\right)_, \label{19}
\end{equation}
and doing a similarity transformation on Eq.(\ref{16}), we have
\begin{equation}
\tilde{H}_D \tilde{\psi}(x)={\cal E} \tilde{\psi}(x), \label{20}
\end{equation}
where
\begin{equation}
\tilde{H}_D=UH_DU^{-1}=\left(\begin{array}{cc} 0 &
-\frac{d}{dx}+me^{\frac{1}{2}\sigma(x)}
\\
\\ \frac{d}{dx}+me^{\frac{1}{2}\sigma(x)}  &
0
\end{array}\right)_, \label{21}
\end{equation}
and
\begin{equation}
\tilde{\psi}(x)=U\psi(x)=e^{\frac{1}{4}\sigma(x)}\left(\begin{array}{c}
\psi_1(x) \\ \psi_2(x)
\end{array}\right)_.
\label{22}
\end{equation}
Now, we show that the Dirac equation (\ref{20}) is equivalent to
the spinor field equation which is obtained from the
(1+1)-dimensional Lagrangian in the flat space-time
\begin{equation}
{\cal L}=i\bar{\tilde{\Psi}}\gamma^a\partial_a
\tilde{\Psi}-\bar{\tilde{\Psi}}\tilde{\Psi} W, \label{23}
\end{equation}
where $W(x)$ is a static scalar function to be determined later.

The field equation derived from the Lagrangian (\ref{23}) becomes
\begin{equation}
i\gamma^a\partial_a\tilde{\Psi}(t, x)-W(x)\tilde{\Psi}(t, x)=0.
\label{25}
\end{equation}
By using the representations (\ref{8}), and considering the time
dependence of the spinor as $\tilde{\Psi}(t, x)=e^{-i{\cal
E}t}\tilde{\psi}(x)$, the equation of motion (\ref{25}) after
multiplying by $\gamma^0$ from left, can be written as
\begin{equation}
\left(\begin{array}{cc} 0 & -\frac{d}{dx}+W(x)
\\
\\ \frac{d}{dx}+W(x)  &
0
\end{array}\right)\tilde{\psi}(x)={\cal E}\tilde{\psi}(x). \label{26}
\end{equation}
Now, considering Eqs.(\ref{26}) and (\ref{20}), it is seen that
they have equivalent mathematical structures. The function $W(x)$
in Eq.(\ref{26}) is commonly called {\it superpotential} \cite{6',
7'} in the context of supersymmetric quantum mechanics. By direct
comparison of (\ref{26}) and (\ref{20}) the conformal factor in
the metric (\ref{10}) is related to the superpotential $W(x)$
through the relation
\begin{equation}
W(x)=me^{\frac{1}{2}\sigma(x)}.
\label{27}
\end{equation}
Now we define the operators ${\cal A}^\dagger$ and ${\cal A}$,
respectively as
\begin{equation}
 {\cal A}^\dagger
:=-\frac{d}{dx}+W(x),
\end{equation}
\begin{equation}
{\cal A} :=\frac{d}{dx}+W(x).
\end{equation}
Therefore, Eq.(\ref{26}) can be written as
\begin{equation}
\left(\begin{array}{cc} 0 & {\cal A}^\dagger
\\
\\ {\cal A}  &
0
\end{array}\right)\tilde{\psi}(x)={\cal E}\tilde{\psi}(x). \label{29}
\end{equation}
By operating $\tilde{H}_D$, defined by Eq.(\ref{21}), from left on
both sides of Eq.(\ref{20}) ( or Eq.(\ref{29}) ) we obtain
\begin{equation}
H_-\tilde{\psi}_1(x)={\cal E}^2\tilde{\psi}_1(x), \label{30}
\end{equation}
\begin{equation}
H_+\tilde{\psi}_2(x)={\cal E}^2\tilde{\psi}_2(x), \label{31}
\end{equation}
where $H_-={\cal A}^\dagger {\cal A}$ and $H_+={\cal A}{\cal
A}^\dagger$ are {\em Partner Hamiltonians} \cite{6', 7'}.
Equations (\ref{30}) and (\ref{31}) can now be solved by
supersymmetric quantum mechanical methods for shape invariant
potentials.

For a typical example we study the superpotential associated with
the modified harmonic oscillator, namely \cite{6'}
\begin{equation}
W(x)=\frac{1}{2}\omega |x|+ c \label{32},
\end{equation}
where $\omega$ and $c$ are real positive quantities. Using
equations (\ref{27}) and (\ref{32}) the metric (\ref{10}) is
obtained
\begin{equation}
ds^2=\left(\frac{\omega|x|}{2m}+\frac{c}{m}\right)^2 (dt^2-dx^2).
\label{33}
\end{equation}
Considering the superpotential (\ref{32}), the partner
Hamiltonians $H_{\pm}$ are
$$
\left \{ \begin{array}{ll}
H_+=-\frac{d^2}{dx^2}+\frac{1}{4}\omega^2x^2+\frac{1}{2}\omega+\omega x c+c^2 \hspace{15mm}\\
\\
H_-=-\frac{d^2}{dx^2}+\frac{1}{4}\omega^2x^2-\frac{1}{2}\omega+\omega
x c+c^2 \hspace{11mm}
\end{array}\right. \vspace{10mm} x>0,
$$
\begin{equation}
\left \{ \begin{array}{ll}
H_+=-\frac{d^2}{dx^2}+\frac{1}{4}\omega^2x^2-\frac{1}{2}\omega-\omega x c+c^2 \hspace{15mm}\\
\\
H_-=-\frac{d^2}{dx^2}+\frac{1}{4}\omega^2x^2+\frac{1}{2}\omega-\omega
x c+c^2 \hspace{11mm}
\end{array}\right. \vspace{10mm} x<0.
\label{34}
\end{equation}
Now, by inserting
$\tilde{\psi}_i(x)=e^{-\frac{\omega}{4}(x+\frac{2c}{\omega})^2}
\phi_i(x)$  $(i=1, 2)$ in Eqs.(\ref{30}), (\ref{31}) and change of
variable $y=\sqrt{\frac{\omega}{2}}(x+\frac{2c}{\omega})$ for
$x>0$, we find the following differential equations
$$
\frac{d^2\phi_1}{dy^2}-2y\frac{d\phi_1}{dy}+2\frac{{\cal
E}^2}{\omega} \phi_1=0,
$$
\begin{equation}
\frac{d^2\phi_2}{dy^2}-2y\frac{d\phi_2}{dy}+2 [\frac{{\cal
E}^2}{\omega}-1]  \phi_2=0,
\end{equation}
respectively. In the same way, inserting
$\tilde{\psi}_i(x)=e^{-\frac{\omega}{4}(-x+\frac{2c}{\omega})^2}
\phi_i(x)$  $(i=1, 2)$ and change of variable
$y=\sqrt{\frac{\omega}{2}}(-x+\frac{2c}{\omega})$ for $x<0$ leads
to
$$
\frac{d^2\phi_1}{dy^2}-2y\frac{d\phi_1}{dy}+2 [\frac{{\cal
E}^2}{\omega}-1]  \phi_1=0,
$$
\begin{equation}
\frac{d^2 \phi_2}{dy^2}-2y\frac{d\phi_2}{dy}+2\frac{{\cal
E}^2}{\omega} \phi_2=0,
\end{equation}
respectively. The wave functions $\tilde{\psi}_1(x)$ and
$\tilde{\psi}_2(x)$ in Eqs.(\ref{30}) and (\ref{31}) are given by
\begin{equation}
\tilde{\psi}_1(x)=\left \{ \begin{array}{ll}
e^{-\frac{\omega}{4}(x+\frac{2c}{\omega})^2}H_n(\sqrt{\frac{\omega}{2}}(x+\frac{2c}{\omega})) \hspace{35mm}x>0\\
e^{-\frac{\omega}{4}(-x+\frac{2c}{\omega})^2}H_{n-1}(\sqrt{\frac{\omega}{2}}(-x+\frac{2c}{\omega}))
\hspace{26mm}x<0,
\end{array}\right.
\label{36}
\end{equation}
\begin{equation}
\tilde{\psi}_2(x)=\left \{ \begin{array}{ll}
e^{-\frac{\omega}{4}(x+\frac{2c}{\omega})^2}H_{(n-1)}(\sqrt{\frac{\omega}{2}}(x+\frac{2c}{\omega})) \hspace{31mm}x>0\\
e^{-\frac{\omega}{4}(-x+\frac{2c}{\omega})^2}H_n(\sqrt{\frac{\omega}{2}}(-x+\frac{2c}{\omega}))
\hspace{32mm}x<0.
\end{array}\right.
\label{37}
\end{equation}
where $H_n$ and $H_{n-1}$ are Hermite polynomials leading to the
following energy spectrum
\begin{equation}
{\cal E}_n=\pm\sqrt{n \omega}, \label{35}
\end{equation}
for both $x<0$ and $x>0$, where $n$ is a non-negative integer
number. Now, using Eqs.(\ref{15}), (\ref{22}), (\ref{36}), and
(\ref{37}), the final solution of Dirac equation is obtained
\begin{equation}
\Psi_n(X)=\left \{ \begin{array}{ll}
a(\frac{\omega
x}{2m}+\frac{c}{m})^{-\frac{1}{2}}e^{-\frac{1}{4}\omega (x+\frac{2c}{\omega})^2-i\sqrt{n\omega}t}\left(\begin{array}{c} H_n(\sqrt{\frac{\omega}{2}}(x+\frac{2c}{\omega}))\\
H_{n-1}(\sqrt{\frac{\omega}{2}}(x+\frac{2c}{\omega}))\end{array}\right)+C.C \hspace{15mm}x>0,\\
\\
b(\frac{-\omega
x}{2m}+\frac{c}{m})^{-\frac{1}{2}}e^{-\frac{1}{4}\omega (-x+\frac{2c}{\omega})^2-i\sqrt{n\omega}t}\left(\begin{array}{c} H_{n-1}(\sqrt{\frac{\omega}{2}}(-x+\frac{2c}{\omega}))\\
H_{n}(\sqrt{\frac{\omega}{2}}(-x+\frac{2c}{\omega}))\end{array}\right)+C.C \hspace{8mm}x<0,\\
\end{array}\right.
\label{38}
\end{equation}
where $a$ and $b$ are complex constants and $C.C$ means complex
conjugation.

We note that the ansatz (\ref{15}) does not work for $n=0$ quantum
number. Therefore, we restrict ourselves to the positive integer
for the quantum numbers, $n$. This gives rise to a broken
supersymmetry \cite{6'}. The continuity condition for the spinor
wave function (\ref{38}) at $x=0$ implies
\begin{equation}
\left \{ \begin{array}{ll}
a H_n(\lambda)=bH_{n-1}(\lambda)\\
a H_{n-1}(\lambda)=bH_n(\lambda)\\
a^* H_n(\lambda)=b^*H_{n-1}(\lambda)\\
a^* H_{n-1}(\lambda)=b^*H_n(\lambda),\\
\end{array}\right.
\label{95}
\end{equation}
where the positive constant $\lambda$ is defined as follows
\begin{equation}
\lambda=c\sqrt{\frac{2}{\omega}}.
\end{equation}
Eqs.(\ref{95}) have non-trivial solutions if
\begin{equation}
\left \{ \begin{array}{ll}
H_n(\lambda)=H_{n-1}(\lambda)) \\
\\
or\\
\\
H_n(\lambda)=-H_{n-1}(\lambda))_.
\end{array}\right.
\label{40}
\end{equation}
For example when $n=1$, according to Eqs.(\ref{95}), the allowed
value for $\lambda$ is $\frac{1}{2}$ (
$c=\frac{1}{2}\sqrt{\frac{\omega}{2}}$ ), and for $n=2$ the
corresponding allowed values are  $\lambda=1$ (
$c=\sqrt{\frac{\omega}{2}}$ ) and $\lambda=\frac{1}{2}$ (
$c=\frac{1}{2}\sqrt{\frac{\omega}{2}}$ ). Therefore, the
continuity of spinor wave function (\ref{38}) at $x=0$ requires,
for each mode $n$, a relation between the constants $\omega$ and
$c$ in Eq.(\ref{32}).

\section*{Acknowledgment}

The authors would like to thank the referee for useful comments.
This work has been supported by the Research Institute for
Fundamental Sciences, Tabriz, Iran.


\begin{thebibliography}{99}
\bibitem{1}J. F. Donoghue and B. R. Holstein, Am. J. Phys. 54, 827
(1986).
\bibitem{2}D. G. Boulware, Phys. Rev. D{\bf 11}, 1404 (1975),
Phys. Rev. D{\bf 12}, 350 (1975).
\bibitem{3}D. N. Page, Phys. Rev. D{\bf 16}, 2402 (1977).
\bibitem{4}M. Martellini, Phys. Rev. D{\bf 16}, 3418 (1977).
\bibitem{5}R. B. Mann, S. Morsink, A. E. Sikkema and T. G.
Steele, Phys. Rev. D{\bf 43}, 3948 (1991).
\bibitem{8'}S. K. Moayedi and F. Darabi, J. Math. Phys. {\bf 42},
1229 (2001).
\bibitem{5'}F. Cooper, A. Khare, R. Musto and A. Wipf, Ann. Phys.
{\bf 187}, 1 (1988).
\bibitem{6'}F. Cooper, A. Khare and U. Sukhatme, Phys. Rep. {\bf
251}, 267 (1995).
\bibitem{1'}S. Weinberg, {\em Gravitation and Cosmology} ( Wiley,
New York, 1972 ).
\bibitem{2'}R. A. Bertlmann, {\em Anomalies in Quantum Field
Theory} ( Oxford University press, New York, 2000 ).
\bibitem{4'}R. Jackiw and C. Rebbi, Phys. Rev. D{\bf 13}, 3358
(1976).
\bibitem{7'}G. Junker, {\em Supersymmetric Methods in Quantum and
Statistical Physics} (Springer-Verlag, Berlin 1996).
\end{thebibliography}
\end{document}